\documentclass[prb,twocolumn,amsmath,amssymb,showpacs]{revtex4-1}
\usepackage{graphicx}

\def\beq{\begin{eqnarray}}
\def\eeq{\end{eqnarray}}
\def\beqa{\begin{eqnarray}}
\def\eeqa{\end{eqnarray}}

\begin{document}

\title{Two distinct quasiparticle inelastic scattering rates in  the $t-J$ model and their relevance for high-$T_c$ cuprates superconductors}

\author{Guillermo Buzon and  Andr\'es Greco}
\affiliation{
Facultad de Ciencias Exactas, Ingenier\'{\i}a y Agrimensura and
Instituto de F\'{\i}sica Rosario
(UNR-CONICET).
Av. Pellegrini 250-2000 Rosario-Argentina.
}

\date{\today}

\begin{abstract}
The recent findings about two distinct quasiparticle inelastic scattering rates in angle-dependent magnetoresistance (ADMR) experiments in
overdoped high-$T_c$ cuprates superconductors have motivated many discussions related to the link between superconductivity, pseudogap, and 
transport properties in these materials. After computing dynamical  self-energy corrections in the framework of the 
$t-J$ model the inelastic scattering rate was introduced as usual. Two distinct scattering rates were obtained showing the main features observed in ADMR experiments.
Predictions for underdoped cuprates are discussed. 
The implicances of these two scattering rates on the resistivity were also studied as a function of doping and temperature and confronted with 
experimental measurements. 
\end{abstract}


\maketitle

\section{Introduction}

The way in which superconductivity (SC) occurs in high-$T_c$  cuprates challenges old concepts in 
condensed matter physics. The puzzling doping dome-shape for the superconducting critical temperature ($T_c$) is tied to the 
anomalous properties observed in the normal state.  In underdoped (UD) the decay of $T_c$ with decreasing doping is correlated to 
the increasing  pseudogap (PG) feature observed above $T_c$. The  so-called PG phase shows properties which are not expected to occur in 
Fermi liquids.\cite{timusk99} 
The anomalous properties observed in UD weaken with increasing doping towards overdoped (OD); however
whether the conventional Fermi liquid (FL) applies in this doping region is controversial. Recent angle-dependent magnetoresistance (ADMR) experiments 
in OD Tl2201 brought insights to this discussion.\cite{abdel06,abdel07,french09} 
These experiments, differently to the resistivity, have the advantage to separate 
the scattering rate ($1/\tau$) in two distinct components. While one component is isotropic ($1/\tau_{i}$) on the Fermi surface (FS) 
and rather constant with doping, the other one is strongly anisotropic ($1/\tau_{a}$) and resembles the anisotropy 
of the PG showing maximum values near the antinode. Moreover, $1/\tau_a$ decreases with increasing doping and vanishes
in highly OD samples. These and others experiments have been recently interpreted (see Ref.[\onlinecite{taillefer10}] 
and references therein) in terms of that the PG phase is distint to SC, ended at the quatum critical point (QCP),  
and its fluctuations as the responsible for pairing and transport properties. 
The present paper shows that two scattering rates, with similar characteristics to those observed in ADMR experiments, are obtained from the $t-J$ model. 

The paper is organized as follows. In Sec. II, a summary of the method is given. The relationship between dynamical self-enegry 
contributions and the two inelastic scattering rates is shown. Sec. III contains the results. In subsection A results for high 
doping values are presented and compared with available experiments in this doping region. In subsection B predictions for low doping are studied and discussed.
In subsection C the implications of the two obtained scattering rates on the resistivity are studied and confronted with measurements. 
Discussion and conclusion are given in Sec. IV.

\section{Summary of the formalism: Inelastic scattering rate}

In leading order of large $N$ expansion ($N$ is the number of spin components) the $t-J$ model predicts a phase diagram with close similarities to 
the phase diagram of hole doped cuprates.\cite{cappelluti99} 
The pseudogap, which is associated to the flux-phase (FP),\cite{affleck88} competes and coexists with SC. 
The large-$N$ mean-field treatment of the $t-J$ model yields 
a quasiparticle (QP) dispersion 

\begin{eqnarray}
\epsilon_{k}=&-&2(t \delta+rJ)(cos(k_x)+cos(k_y))\\ \nonumber
&+&4t' \delta cos(k_x) cos(k_y)-\mu
\end{eqnarray}
where the parameters $t$, $t'$ and $J$ are the hopping between nearest-neighbor, next nearest-neighbor and the nearest-neighbor 
Heisenberg coupling respectively. The contribution  
$r$ to the mean-field band and the chemical potential $\mu$ must be obtained self-consistently\cite{greco04} from  

\begin{eqnarray}
r=\frac{1}{N_s} \sum_{\bf k} cos(k_x) n_F(\epsilon_{k})
\end{eqnarray}
and

\begin{eqnarray}
(1-\delta)=\frac{1}{N_s} \sum_{\bf k} n_F(\epsilon_{k})
\end{eqnarray}
where $n_F$ is the Fermi function, $\delta$ the doping away from half-filling and, $N_s$ the number of sites. 

Hereafter, $t'/t=0.35$, $J/t=0.3$, suitable for cuprates, are used and  
the lattice constant $a$ of the square lattice is considered as length unit.

In this approach, the mean-field homogeneous state becomes unstable when the static ($\omega=0$) flux susceptibility 

\begin{eqnarray}
\chi_{flux}({\bf q},\omega)=(\frac{\delta}{2})^2 [(8/J) r^2-\Pi({\bf q},\omega)]^{-1}
\end{eqnarray}
diverges.\cite{greco08} In Eq.(4) $\Pi({\bf q},\omega)$ 
is an electronic polarizability  calculated with a form factor 
$\gamma({\bf q},{\bf k})=2 r (sin(k_x-q_x/2)-sin(k_y-q_y/2))$.

In Fig.5(b), disregarding SC, the solid line shows, in the doping-temperature ($\delta-T$) plane, the temperature 
$T_{FP}$ which indicates the onset of FP 
instability. At $T_{FP}$ a flux-mode ($Im \chi_{flux}({\bf q}=(\pi,\pi),\omega)$) reaches $\omega=0$ freezing the FP.\cite{greco08} At $T=0$ a phase transition occurs 
at the QCP placed at the critical  doping $\delta_c \cong 0.17$ (Fig.5(b)). Since the instability takes place at  
$(\pi,\pi)$ the form factor $\gamma({\bf q},{\bf k})$ transforms 
into $\sim (cos(k_x)-cos(k_y))$ which indicates the $d$-wave character of the FP. Although the relevance of the FP for the physical case $N=2$, for instance in the form of a phase without long-range but strong $d$-wave 
short-range order, is under dispute,\cite{leung00} the FP scenario possesses the main properties to be identified with the phenomenological 
$d$ CDW  proposal.\cite{chakravarty01} 

For discussing the quasiparticle inelastic scattering rate it is necessary to calculate self-energy corrections 
collecting ${\cal O} (1/N)$ fluctuations
beyond mean field. As showed in Ref.[\onlinecite{bejas06}] the self-energy $\Sigma({\mathbf{k}},\omega)$ contains contributions from 
six different channels and their mixing: The usual charge channel named $\delta R$, a non-double-occupancy channel named $\delta \lambda$ and, four charge channels driven by $J$. 
However, as discussed in Ref.[\onlinecite{greco08}] and summarized below, the 
relevant contributions to $\Sigma({\mathbf{k}},\omega)$ can be written as

\begin{eqnarray}\label{eq:SigmaIm0}
Im \, \Sigma({\mathbf{k}},\omega)=Im \, \Sigma_{R \lambda}({\mathbf{k}},\omega)+Im \, \Sigma_{flux}({\mathbf{k}},\omega)
\end{eqnarray}
\noindent where 
\begin{eqnarray}\label{eq:SigmaIm0}
Im \, \Sigma_{R \lambda}({\mathbf{k}},\omega)&=&-\frac{1}{ N_{s}}
\sum_{{\mathbf{q}}} \left\{ \Omega^{2} \;
 Im[D_{RR}({\mathbf{q}},\omega-\epsilon_{{k-q}})] \right. \nonumber\\
&& \hspace{-2.5cm} + \; 2\;\Omega \; Im[D_{\lambda R}({\mathbf{q}},\omega-\epsilon_{{k-q}})]
+ \left. Im[_{\lambda \lambda}({\mathbf{q}},\omega-\epsilon_{{k-q}})] \right\} \nonumber\\
&& \hspace{-1cm}\times \left[n_{F}(-\epsilon_{{k-q}}) + n_{B}(\omega-\epsilon_{{k-q}})\right],
\end{eqnarray}
\noindent and 
\begin{eqnarray}\label{eq:SigmaIm0}
Im \, \Sigma_{flux}({\mathbf{k}},\omega)&=&-\frac{1}{N_{s}}\sum_{{\mathbf{q}}} \gamma^2({\bf q},{\bf {k}}) Im \chi_{flux}({\bf q},\omega-\epsilon_{{k-q}}) \nonumber\\
&& \hspace{-1cm}\times \left[n_{F}(-\epsilon_{{k-q}}) + n_{B}(\omega-\epsilon_{{k-q}})\right]
\end{eqnarray}
In the above expressions, $\Omega=(\varepsilon_{{k-q}}+\omega+\mu)/2$ and $n_B$ is the Bose function.
 
The physical meaning of Eq.(5) is as  follows. 
$\Sigma_{R \lambda}$ corresponds to 
the usual charge ($\delta R$) sector,  non-double-occupancy  ($\delta \lambda$) sector and the mixing of both. 
For $J/t=0.3$ there is no important  influence of $J$-contributions in 
$\Sigma_{R \lambda}$ showing that the usual charge sector is weakly coupled to the $J$-channels. On the other hand, 
$J$-channels play an important role at low energy and low doping in the proximity to the flux phase instability. 
$\Sigma_{flux}({\mathbf{k}},\omega)$ (Eq.(7)) is obtained after projecting the self-energy on the eigenvector corresponding to the flux instability. Eq.(7) shows 
the coupling between carriers and FP fluctuations.  
For the explicit expressions of $D_{RR}$, $D_{\lambda R}$ and $D_{\lambda \lambda}$ see 
Ref.[\onlinecite{bejas06}]. 

Motivated from above discussion and the results presented  below the two scattering rates $1/\tau_i$ and $1/\tau_a$ (discussed in the 
introduction) are associated to $\Sigma_{R \lambda}$ and $\Sigma_{flux}$, respectively, as follows

\begin{eqnarray}
1/\tau_{a}({\mathbf{k_F}}) \equiv -2 Im \Sigma_{flux}({\mathbf{k_F}},\omega=0)
\end{eqnarray}
 
\begin{eqnarray}
1/\tau_{i}({\mathbf{k_F}}) \equiv -2 Im \Sigma_{R\lambda}({\mathbf{k_F}},\omega=0)
\end{eqnarray}
where ${\mathbf{k_F}}$ is a momentum on the FS.

\section{Results}

\subsection{Overdoped results: Comparison with ADMR experiments}

Fig.1 shows, for two dopings and different temperatures, $1/\tau_{a}({\mathbf{k_F}})$ and  $1/\tau_{i}({\mathbf{k_F}})$ on the FS labeled by the 
angle $\phi$ ranging from the antinode ($\phi=0$)  to the node ($\phi=\pi/4$) (see inset in panel (b2)). 
Note that $\delta=0.20$ and $\delta=0.25$ lay in OD as the samples studied in ADMR experiments. 
Similarly to the behavior observed in ADMR\cite{abdel06,abdel07} 
$1/\tau_{i}$ is very isotropic on the FS and  $1/\tau_{a}$ is strongly anisotropic showing the maximum near the antinode. 
Note that $1/\tau_{a}$ follows, approximately, the proposed\cite{abdel06,abdel07} shape $1/\tau_a \sim cos^2(2\phi)$ (dashed line in 
panels (b1) and (b2)). 
Ossadnik {\it et al.} (Ref.[\onlinecite{ossadnik08}]) have shown similar results for moderate onsite Coulomb repulsion on the Hubbard model 
in one loop renormalization group approximation. 
\begin{figure}
\begin{center}
\setlength{\unitlength}{1cm}
\includegraphics[width=8cm,angle=0.]{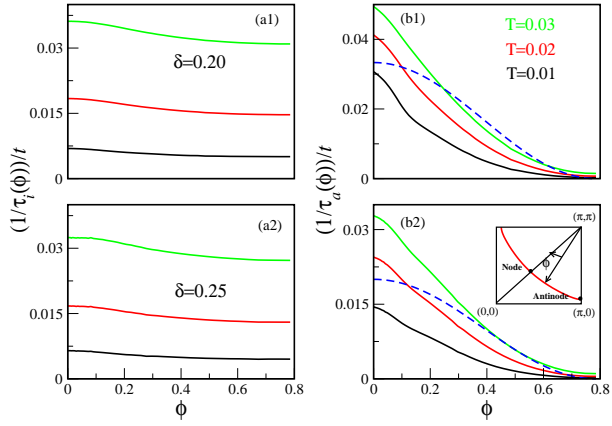}
\end{center}
\caption{(Color online) (a1) and (a2) isotropic scattering rate $1/\tau_{i}$ on the Fermi surface  for $\delta=0.20$ and $\delta=0.25$ 
respectively, and for $T/t=0.03,0.02$ and $0.01$. (b1) and (b2) the same than panels (a1) and (a2)
for the anisotropic scattering rate $1/\tau_{a}$. Dashed line in (b1) and (b2)
is a guide for the eyes $\sim cos^2(2\phi)$ showing a similar trend for
$1/\tau_a$. Inset in (b2) defines the angle $\phi$ ranging for the antinode to the node. 
In the calculation of $\chi_{flux}$, $\eta=0.02t$ was used in the analytical continuation.
}
\end{figure}

In Fig.2(a), $<1/\tau_{a}>$ and $<1/\tau_{i}>$ (where $<>$ means the average on the FS)
are plotted as a function of doping for a fixed temperature $T/t=0.01$. This temperature is  close 
to the reported $T\sim 50K$ in Fig.2 of Ref.[\onlinecite{abdel07}]  if the accepted value $t=0.4eV$ is considered.
As in the experiments, $<1/\tau_{i}>$ is only weakly doping dependent and,
$<1/\tau_{a}>$ is strongly decreasing with increasing doping. 
Inset of Fig.2(a) shows, in the $\delta-T$ plane, that $1/\tau_{a}$ dominates over $1/\tau_{i}$ at low  doping and low 
temperatures (blue region) while, $1/\tau_a$ dominates on the red region. In Fig.2(b), the ratio $(<1/\tau_{a}>/<1/\tau_{i}>)$ at $T/t=0.01$ is plotted as a function of doping (solid line) 
and compared with the experimental data (solid circles) taken from Ref.[\onlinecite{abdel07}]. Although the experimental data  
seem to decay faster than in the theory, both results show a similar trend with increasing doping. 
Note that while theory predicts that $1/\tau_{a}$ 
becomes smaller than $1/\tau_{i}$ for $\delta \sim 0.29$ (see also Fig.2 (a)), the same occurs in the experiment for $\delta \sim 0.26$. 
Beyond a quantitative comparison it is important to  discuss about the physical interpretation. 
In Refs.[\onlinecite{abdel07,taillefer10}] it was claimed that $1/\tau_{a}$ vanishes just at the doping value where SC 
emerges  from OD following dashed line in Fig.2(b), concluding  that the source for SC and for the anisotropic scattering rate is the same 
and associated to PG fluctuations.
In present case $1/\tau_{a}$ is caused by the scattering between carriers and short-range FP fluctuations. Thus, since 
$\Sigma_{flux}$ proves the proximity to the PG via the coupling between carriers and the soft flux-mode of momentum $(\pi,\pi)$, 
$1/\tau_a$ is strongly anisotropic on the FS and decreases with increasing doping beyond the QCP. However, SC 
comes from the instantaneous (no bosonic glue) short-range antiferromagnetic exchange $J$\cite{cappelluti99,anderson07} leading to a different origin for SC and the scattering rate. Finally, it is worth to mention that considering 
the error bars (Fig.2(b)) $1/\tau_a$ could also decay following, for instance,  dotted line which is nearly paralel to the theoretical predicted solid line.

\begin{figure}
\begin{center}
\setlength{\unitlength}{1cm}
\includegraphics[width=8cm,angle=0.]{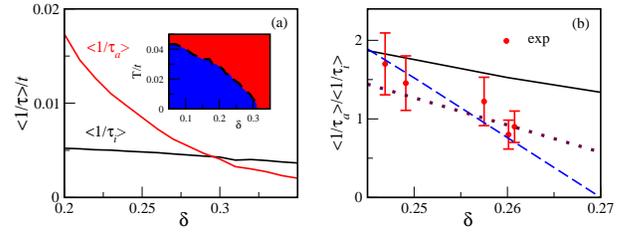}
\end{center}
\caption{(Color online) 
(a) $<1/\tau_i>$ and $<1/\tau_a>$ as a function of doping at $T/t=0.01$.
The inset shows, in the $\delta-T$ plane, that $1/\tau_a$ is larger than
$1/\tau_i$ at low doping and low temperatures (blue region), in the red
region $1/\tau_i$ is larger.
(b) ($<1/\tau_a>/<1/\tau_i>$) ratio at $T/t=0.01$ as a function of doping (solid line).
Solid circles are the experimental data at similar temperature shown for comparison (see text for discussion).
Dashed and dotted lines are possible trends according to the error bars. 
}
\end{figure}

In Fig. 3 the temperature behavior of both scattering rates is shown. At low temperature 
$<1/\tau_{i}> \sim  T^m$ with  $m \sim 2$ (Fig. 3(a)) for all dopings. A similar quadratic temperature behavior for $1/\tau_i$ 
was observed for the isotropic scattering rate in OD 
Tl2201.\cite{abdel06,french09}
In contrast, $<1/\tau_{a}>$ (Fig. 3(b)) shows, at high dopings (see results for $\delta=0.27$ and $\delta=0.30$), a different behavior: 
At low temperature,  $<1/\tau_{a}> \sim T^m$ with $m \sim 1$ 
which is close to the $T$-linear law discussed in Refs.[\onlinecite{abdel06,french09}] for similar dopings.
Therefore, it is  concluded that, besides the anisotroy on the FS,  at high doping the temperature 
dependence for both, $1/\tau_a$ and $1/\tau_i$, agrees also, qualitatively, 
with the experiment.

\begin{figure}
\begin{center}
\setlength{\unitlength}{1cm}
\includegraphics[width=8cm,angle=0.]{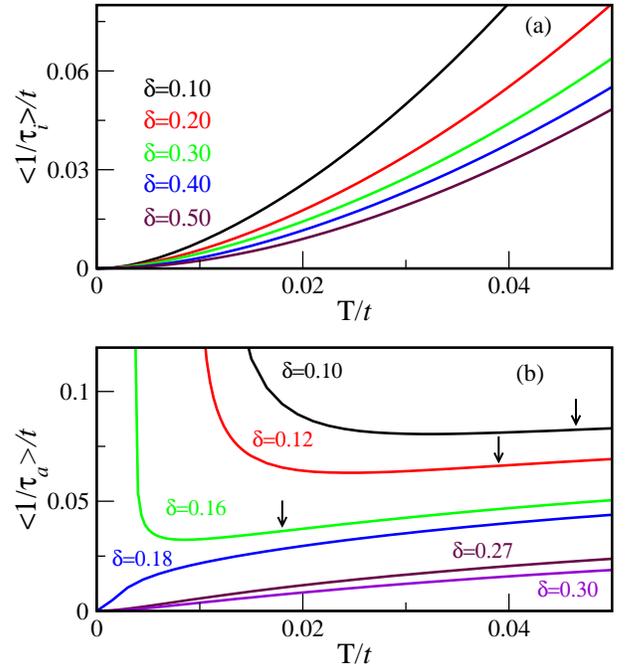}
\end{center}
\caption{(Color online) 
(a) $<1/\tau_i>$ as a function of temperature
for several dopings. 
(b) The same than (a) for $<1/\tau_a>$. 
The studied dopings are indicated in each panel. 
Arrows in (b) indicate the temperature where
an upturn is observed for low dopings (see text).
}
\end{figure}

\subsection{Underdoped results: Possible predictions}

To the knowledge of the authors there are no ADMR experiments in the UD region, however for low doping ($\delta <\delta_c$) the present approach shows predictions which in principle can be tested by ADMR experiments if they 
are possible.
With decreasing temperature $<1/\tau_{a}>$ 
shows an upturn below a given temperature ($T^{up}$) indicated by arrows (see results for $\delta=0.10$, $0.12$ and, $0.16$ in Fig.3(b)). 
$T^{up}$ increases with decreasing doping following the same trend 
(see dotted-dashed line in Fig.5(b))
than $T_{FP}$ . 
Interestingly, although $T^{up}$ marks a smooth crossover and not a true phase transition,   
dotted-dashed line terminates at the QCP.

While for large dopings  $1/\tau_{a}$ decreases with decreasing temperature (panels (b1) and (b2) in Fig.1), for $\delta=0.14$ 
(Fig. 4(b1)) and $\delta=0.15$ (Fig. 4(b2)),
$1/\tau_{a}$ shows a reentrant behavior near the antinode.
It is important to mention that the length from the antinode 
of the re-entrant (marked in Fig. 4(b1) and (b2)) expands with decreasing temperature and doping 
resembling the  behavior of the Fermi-arcs observed in angle-resolved photoemission spectroscopy\cite{norman98} (ARPES), i.e.,
when larger is the re-entrant length shorter the arcs. 
See last section  for further discussion about a possible link between ADMR and ARPES experiments.

Finally, Fig.4 (a1) and (a2) together with results for $\delta=0.10$ in Fig. 3(a) show that the evolution of $1/\tau_i$ from UD to OD 
is rather smooth.
\begin{figure}
\begin{center}
\setlength{\unitlength}{1cm}
\includegraphics[width=8cm,angle=0.]{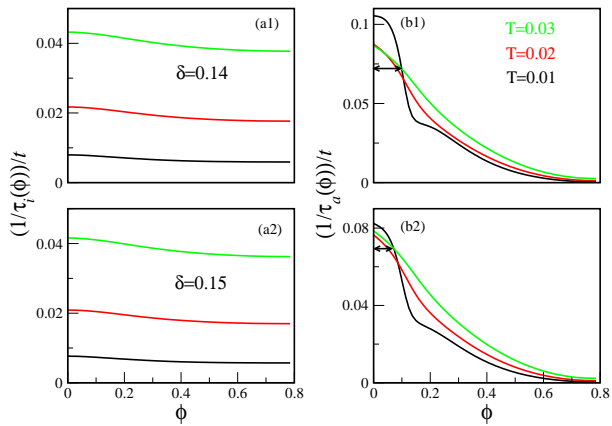}
\end{center}
\caption{(Color online) The same than Fig.1 for the low doping values 
$\delta=0.14$ and $\delta=0.15$. With decreasing temperature a re-entrant behavior occurs near the antinode for $1/\tau_a$ (panels (b1) and (b2)).
}
\end{figure}
\subsection{Resistivity}

In spite of intense studies the origin of the resistivity in cuprates is far from closed. Whether 
the resistivity is  composed by two different contributions with different temperature and doping dependence,\cite{ cooper09} or a single 
contribution is currently under dispute.\cite{hussey08} Therefore, based on present results, it is worth to discuss 
a possible connection between the two scattering rates 
observed in ADMR and resistivity measurements.
In first approximation $\rho=4 \pi / \omega^2_{p} \tau_{tot}$ where $1/\tau_{tot}=<1/\tau_{a}> + <1/\tau_{i}>$
and $\omega_{p}$ is the plasma frequency.  
Using the obtained results for $1/\tau_{tot}$ as a function of temperature (Fig.5(a)) 
a curvature mapping ($\frac{d^2 (1/\tau_{tot})}{dT^2}$) in the $\delta-T$ plane is presented in Fig.5(b) where positive (negative)
curvature is indicated in red (blue). 
It is possible to divide the behavior of $1/\tau_{tot}$ in two regimes. 
(a) $\delta < \delta_c$: 
since at low temperatures $1/\tau_a$ dominates (inset in Fig. 2(a)), an upturn also occurs for $1/\tau_{tot}$ (Fig. 5(a)). Note that 
for $1/\tau_{tot}$ the upturn is shifted to lower temperatures respect to the upturn of the isolated 
$1/\tau_a$ being more pronounced below dashed line in Fig. 5(b). 
A similar upturn for the resistivity was found experimentally.\cite{ando04,daou09}
While Ando {\it et al.}, for BSLCO, LSCO, and YBCO found a weak and negative curvature above the upturn temperature, Daou  {\it et al.},
reported a $T$-linear law for Nd-LSCO. In present case $1/\tau_{tot} \sim  T^m$ with $m \sim 1.5$ indicates a weak positive curvature (light red region) 
closer to the results of Ref.[\onlinecite{daou09}].
At this point it is important to make the following remark. 
As discussed above the upturn is caused by the interaction between quasiparticles and short-range fluctuations of the FP. If long-range order 
sets in
at $T_{FP}$, in the context of present approximation  $1/\tau_{tot}$ is expected to diverge at $T_{FP}$. 
This divergence is due to the fact that FP fluctuations were not self-consistently included  in $\Sigma_{flux}$.\cite{note} The inclusion of these fluctuations would weaken the upturn leading, for 
$T<T_{FP}$, to a $d$ CDW metal. 
(b) $\delta >\delta_c$: close to $\delta_c$ and at low temperature 
$1/\tau_{tot}$ shows a downturn (blue region in Fig. 5(b)) 
which fades out,  with increasing temperature  and doping, faster than in  experiment.\cite{ando04} 
For larger dopings  $1/\tau_{tot}$ shows a positive $T$-curvature.  For low temperature $1/\tau_{tot}$ could be approximated  by a single 
power law  $1/\tau_{tot} \sim  T^m$. With increasing doping $m$ increases from $m\sim 1.2$ near $\delta \sim 0.20$ to $m\sim 2$ at high doping 
(solid line in Fig. 6(a)).
Similar behavior is widely  discussed in the literature \cite{hussey08} and interpreted as the emergence of the FL in OD. 
The experimental values for $m$ (solid circles) taken from Ref.[\onlinecite{kubo91}] were included for comparison,
showing a rather good agreement between theory and experiment (see also Ref.[\onlinecite{naqib03}]).

\begin{figure}
\begin{center}
\setlength{\unitlength}{1cm}
\includegraphics[width=8cm,angle=0.]{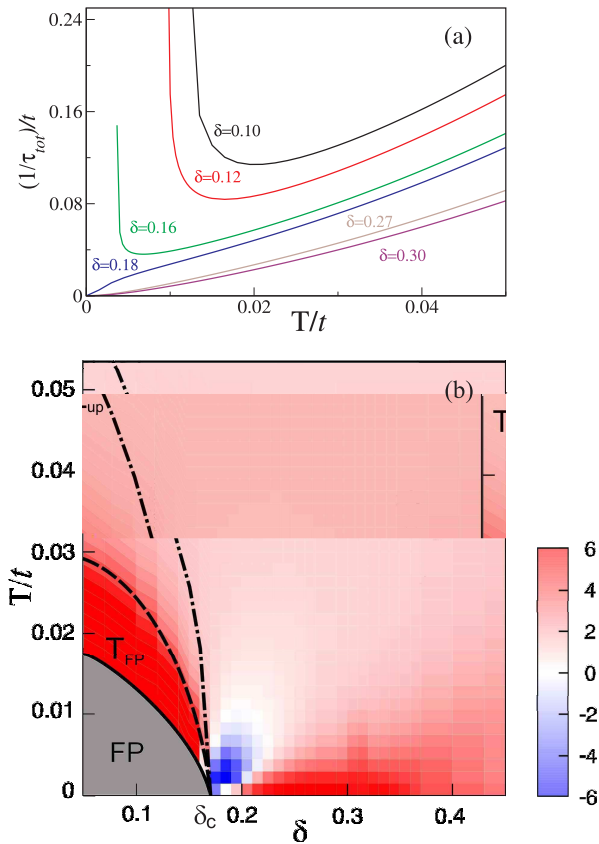}
\end{center}
\caption{(Color online) (a) The same than Fig. 3(a) and Fig. 3(b) for $<1/\tau_{tot}>$.
(b) Curvature mapping ($\frac{d^2 (1/\tau_{tot})}{dT^2}$) in the $\delta-T$
plane. Red (blue) region indicates positive (negative) curvature.
}
\end{figure}

The doping dependence of the resistivity is also important. 
In Fig.6(b) the $log(\rho)$ for  Bi2201\cite{kondo06} (circles)
and for Tl2201\cite{ma06} (diamonds) is plotted as a function of doping
(left axis) together with $log(1/\tau_{tot})$ (right axis) at $T=200K$. 
This figure shows
that both, $\rho$ and $1/\tau_{tot}$, follow similar doping dependence
suggesting that the scattering rate is the main cause for the
doping behavior of the resistivity. Comparing the scales of right and left axis
it is possible to see that they are shifted by a constant $C \sim 8$, since
$C=log(4\pi/\omega_p^2)$, then $\omega_p \sim 1eV$ which is somewhat lower but on the order
of the experiment\cite{timusk99} and also consistent with present dispersion
$\epsilon_{k}$ (Eq.(1)). 

Finally, the resistivity for the high doping case $\delta=0.40$ is estimated. 
Using $\omega_{p}=1eV$, $\rho$ has the following similarities with measurements in high OD  LSCO:\cite{nakamae03} (a) For $T < 50K$, 
$\rho \sim  A T^m$ with $m\sim 2$, while for higher temperatures $m\sim 1.6$. (b) The quadratic coefficient $A$ is 
$A \sim 4 \;n \Omega cm/K^2$ which is on the order of magnitude reported in Ref.[\onlinecite{nakamae03}]. 
This high value of $A$, which is about two orders of magnitude larger than the expected value for conventional metals,  
supports the interpretation that,
even at very high dopings, cuprates must be considered 
in the strong-coupling regime.\cite{nakamae03}

In summary, considering that the experimental behavior of the resistivity is rather controversial\cite{hussey08} and no existing theory describing 
systematically all observed features, present results  can be considered satisfactory.

\begin{figure}
\begin{center}
\setlength{\unitlength}{1cm}
\includegraphics[width=8cm,angle=0.]{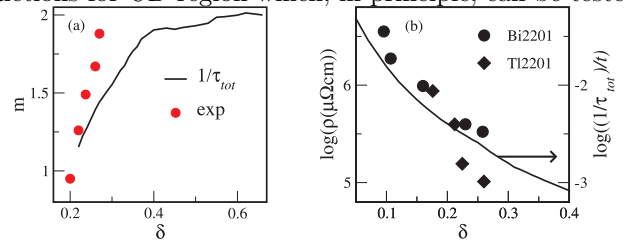}
\end{center}
\caption{(Color online) (a) Estimated doping dependence of the temperature exponent $m$ assuming the approximated form 
$<1/\tau_{tot}> \sim T^m$. Solid circles are
taken from  the experiment for comparison. (b) Logarithm of the experimental resistivity ($log(\rho)$) (left axis) and $log(1/\tau_{tot})$ (right axis)
as a function of doping at $T=200K$. 
}
\end{figure}

\section{Discussion and conclusion}

In the framework of the $t-J$ model in large $N$ approximation, dynamical self-energy 
corrections were computed beyond mean field. The existence of two distinct components for the 
quasiparticle inelastic scattering rate was shown. While one component is very anisotropic on the FS and disappears at high dopings, the other 
component is isotropic on the FS and rather constant with doping. In OD, the two components behave qualitatively similar to those observed in ADMR experiments. 
In addition, predictions for UD region which,
in principle,  can be tested in ADMR experiments were discussed. 

The doping and temperature behavior of the estimated resistivity has similarities with the corresponding 
transport measurements concluding that 
the picture presented here may contribute on discussions about the origin of the resistivity and its possible 
link to the findings observed in ADMR experiments.

It is known that the large-$N$ expansion weakens antiferromagnetic fluctuations respect to charge fluctuations. However, 
it is important to mention that ADMR experiments were performed in the OD region where 
the large-$N$ approach is expected to be more reliable. The observed agreement between experiments and theory is interpreted here 
as indication that, at least, part of the physics is captured by present theory. For instance, 
for high doping present calculation predicts $\rho=A T^2$ as observed in the experiments. Importantly, besides the 
temperature behavior, an unusual high value for the coefficient $A$ is obtained in agreement with measurements. This is interpreted 
as an indication that, even in OD, strong coupling effects occur in cuprates.

Before closing, a possible link between ADMR and ARPES experiments is discussed. It was recently shown that (a) $\Sigma_{flux}$, which proves the proximity to the PG, 
dominates at low doping and low energy leading to Fermi-arcs effects.\cite{greco09}
(b) $\Sigma_{R\lambda}$, which contains non-double-occupancy effects of the 
$t-J$ model, dominates at high energy 
leading to high energy features\cite{foussats08} which resemble the waterfall effects observed in ARPES.\cite{graf07}
Thus, present paper suggests also a possible common origin for the features seen in ADMR and ARPES experiments. 

\noindent{\bf Acknowledgments}

The authors thank to M. Bejas, A. Foussats, and H. Parent for valuable discussions.


\begin{thebibliography}{10}

\bibitem{timusk99}
T. Timusk and B. Statt, Rep. Prog. Phys. {\bf 62}, 61 (1999).

\bibitem{abdel06}
M. Abdel-Jawad {\it et al.}, Nature Phys. {\bf 2}, 821 (2006).

\bibitem{abdel07}
M. Abdel-Jawad {\it et al.}, Phys. Rev. Lett.  {\bf 99}, 107002 (2007).

\bibitem{french09}
M.M. J. French {\it et al}, New J. Phys. {\bf 11}, 055057 (2009).

\bibitem{taillefer10}
L. Taillefer, Annual Review of Condensed Matter Physics {\bf 1}, 51 (2010).

\bibitem{cappelluti99}
E. Cappelluti and R. Zeyher, Phys. Rev. B {\bf 59}, 6475 (1999).

\bibitem{affleck88}
I. Affleck and J.B. Marston, Phys. Rev. B {\bf 37}, 3774 (1988).

\bibitem{greco04}
A. Foussats and A. Greco, Phys. Rev. B {\bf 70}, 205123 (2004).

\bibitem{greco08}
A. Greco, Phys. Rev. B {\bf 77}, 092503 (2008).

\bibitem{leung00}
P. W. Leung, Phys. Rev. B {\bf 62}, R6112 (2000). A. Macridin, M. Jarrell, and Th. Maier, Phys. Rev. B {\bf 70},
113105 (2004).

\bibitem{chakravarty01}
S. Chakravarty, R. B. Laughlin, D.K. Morr, and C. Nayak, Phys. Rev. B {\bf 63}, 094503 (2001).

\bibitem{bejas06}
M. Bejas, A. Greco, and A. Foussats, Phys. Rev. B {\bf 73}, 245104 (2006).

\bibitem{ossadnik08}
M. Ossadnik, C. Honerkamp, T. M. Rice, and M. Sigrist, Phys. Rev. Lett.  {\bf 101}, 256405 (2008).

\bibitem{anderson07} P.W. Anderson, Science {\bf 316}, 1705 (2007).

\bibitem{norman98}
M.R. Norman {\it et al.}, Nature {\bf 392}, 157 (1998).
A. Kanigel {\it et al.}, Nat. Phys. {\bf 2}, 447 (2006).

\bibitem{cooper09}
R.A. Cooper {\it et al.}, Science {\bf 323}, 603 (2009). 

\bibitem{hussey08} N.E. Hussey, J. Phys.: Condens. Matter {\bf 20}, 123201 (2008). 

\bibitem{ando04}
Y. Ando {\it et al.}, Phys. Rev. Lett.  {\bf 93},267001 (2004).

\bibitem{daou09}
R. Daou {\it et al.}, Nature Phys. {\bf 5}, 31 (2009).

\bibitem{note} In ${\cal O} (1/N)$, $\Sigma_{flux}$ (Eq.(7)) proves the instability only via the flux-mode which reaches $\omega=0$ at $T_{FP}$. 
In this order, $\Sigma_{flux}$ does not contain, self-consistently,  self-energy effects which are beyond the scope of $1/N$-expansion. 
As discussed in text these effects are expected to play a role only near $T_{FP}$.

\bibitem{kubo91} 
Y. Kubo, Y. Shimakawa, T. Manako, and H. Igarashi, Phys. Rev. B {\bf 43}, 7875 (1991).

\bibitem{naqib03}
S.H. Naqib, J.R. Cooper, J.L. Tallon, and C. Panagopoulos, Phys. C {\bf 387}, 365 (2003).

\bibitem{kondo06}
T. Kondo, T. Takeuchi, S. Tsuda, and S. Shim, Phys. Rev. B {\bf 74}, 224511 (2006).

\bibitem{ma06}
Y.C. Ma and N.L. Wang, Phys. Rev. B {\bf 73}, 144503 (2006)

\bibitem{nakamae03}
S. Nakamae {\it et al.}, Phys. Rev. B {\bf 68}, 100502 (2003)

\bibitem{greco09}
A. Greco, Phys. Rev. Lett.  {\bf 103}, 217001 (2009).

\bibitem{foussats08}
A. Foussats, A. Greco, and M. Bejas,  Phys. Rev. B {\bf 78}, 153110 (2008)

\bibitem{graf07}
J. Graf {\it et al.}, Phys. Rev. Lett. {\bf 98}, 067004 (2007). W. Zhang {\it el al.}, 
Phys. Rev. Lett. {\bf 101}, 017002 (2008).

\end{thebibliography}
\end{document}